\documentclass[aps,prd,twocolumn,superscriptaddress,longbibliography,nofootinbib]{revtex4-2}
\usepackage[utf8]{inputenc}
\usepackage{color}
\usepackage{graphicx}
\usepackage{subfigure}
\usepackage{multirow}

\usepackage{amsmath,amssymb,amsfonts}

\def\prl{Phys. Rev. Lett.}
\def\prd{Phys. Rev. D}

\def\pau_p{Prog. Theor. Phys.}

\def\order{{\mathcal O}}
\def\eq{Eq.}

\begin{document}

\title{Shock-avoiding slicing conditions: tests and calibrations}

\author{Thomas W.~Baumgarte}

\affiliation{Department of Physics and Astronomy, Bowdoin College, Brunswick, ME 04011, USA}

\author{David Hilditch}
\affiliation{
  Centro de Astrof\'{\i}sica e Gravita\c c\~ao -- CENTRA,
  Departamento de F\'{\i}sica, Instituto Superior T\'ecnico -- IST,
  Universidade de Lisboa -- UL, Av.\ Rovisco Pais 1, 1049-001 Lisboa,
  Portugal}

\begin{abstract}
While the 1+log slicing condition has been extremely successful in numerous numerical relativity simulations, it is also known to develop ``gauge-shocks" in some examples. Alternative ``shock-avoiding" slicing conditions suggested by Alcubierre prevent these pathologies in those examples, but have not yet been explored and tested very broadly.  In this paper we compare the performance of shock-avoiding slicing conditions with those of 1+log slicing for a number of ``text-book" problems, including black holes and relativistic stars.  While, in some simulations, the shock-avoiding slicing conditions feature some unusual properties and lead to more ``gauge-dynamics" than the 1+log slicing condition, we find that they perform quite similarly in terms of stability and accuracy, and hence provide a very viable alternative to 1+log slicing.
\end{abstract}

\maketitle

\section{Introduction}
\label{sec:intro}

Numerous applications in numerical relativity adopt the ``Bona-Masso" slicing condition
\begin{equation} \label{bona_masso}
    (\partial_t - \beta^i \partial_i) \, \alpha = - \alpha^2 f(\alpha) K,
\end{equation}
where $\alpha$ is the lapse function, $\beta^i$ the shift vector, and $K$ the mean curvature, i.e.~the trace of the extrinsic curvature (see \cite{BonMSS95}).  A specific slicing condition is then determined by choosing a specific ``Bona-Masso" function $f(\alpha)$; a simple example is $f(\alpha) = 1$, which results in harmonic slicing.  A very common choice in numerical relativity is the ``1+log" condition
\begin{equation} \label{1+log}
    f(\alpha) = \frac{2}{\alpha},
\end{equation}
which has proven to result in many desirable properties.   Together with a ``Gamma-driver" condition for the shift (e.g.~\cite{AlcB01,AlcBDKPST03}), 1+log slicing forms what are sometimes called ``moving-puncture" or ``standard" gauge conditions, which have been used very successfully, for example, in simulations of black-hole binaries (see, e.g., \cite{CamLMZ06,BakCCKM06}).

However, in some applications 1+log slicing can also lead to so-called ``gauge-shocks".  Specifically, the lapse function may develop discontinuities, which are then very difficult to handle numerically (see, e.g., \cite{Alc97,AlcM98} and Section \ref{sec:gaugeshocks} below for examples, as well as \cite{Alc05} for a careful analysis of these shocks).  Alcubierre (see \cite{Alc97,Alc03}) therefore suggested an alternative ``shock-avoiding" choice for the Bona-Masso function, namely
\begin{equation} \label{sa}
    f(\alpha) = 1 + \frac{\kappa}{\alpha^2},
\end{equation}
where $\kappa$ is a constant that we assume to be positive.\footnote{Evidently, the condition (\ref{sa}) reduces to harmonic slicing for $\kappa = 0$.}  While this choice avoids shocks, it also has some unusual properties.  In particular, with $f(\alpha)$ given by (\ref{sa}), the right-hand side of (\ref{bona_masso}) does not vanish for $\alpha = 0$ (and non-zero $K$), so that the lapse may become negative during a time evolution (see the discussion in \cite{Alc03}).  

This observation may help explain why the condition (\ref{sa}) has not been used more widely in numerical relativity applications.  In fact, the only example that we are aware of are simulations of the critical collapse of a non-minimally coupled scalar field.  As reported by \cite{JimVA21}, 1+log slicing (\ref{1+log}) results in gauge pathologies in these simulations that can be avoided by using the shock-avoiding condition (\ref{sa}) instead.  The authors of \cite{Hiletal13} similarly found that using 1+log slicing in simulations of vacuum gravitational waves leads to the development of discontinuities in the lapse function.

While the shock-avoiding slicing condition (\ref{sa}) appears to have noticeable advantages over 1+log slicing in specific examples, it remains unclear how shock-avoiding slicing behaves in other cases, in particular since it may result in negative values for the lapse function.  The purpose of this paper, therefore, is to explore this behavior for a number of  simple ``textbook examples", and compare with that of 1+log slicing.  Specifically, we will, after discussing some algebraic features of shock-avoiding slices as well as our numerical code in Section \ref{sec:prelim}, consider a gauge-pulse problem (Section \ref{sec:gaugeshocks}), Schwarzschild spacetimes (Section \ref{sec:schwarzschild}), Kerr spacetimes (Section \ref{sec:kerr}), the head-on collision of two black holes (Section \ref{sec:head-on}), Oppenheimer-Snyder collapse (Section \ref{sec:os}), and simulations of single neutron stars (Section \ref{sec:tov}).  We conclude with a brief summary in Section \ref{sec:summary}.

Throughout this paper we adopt geometrized units with $G = c = 1$.

\section{Preliminaries}
\label{sec:prelim}

\subsection{Algebraic Expressions}
\label{sec:general}

In many numerical relativity simulations asymptotic flatness implies that, for large distances from any gravitational sources, the value of the lapse function $\alpha$ can be chosen to be unity.  We then start by observing that, for 1+log slicing, the Bona-Masso function (\ref{1+log}) can be expanded about the asymptotic value $\alpha = 1$ to yield
\begin{equation} \label{1+log_f_exp}
    f(\alpha) = 2 + 2 (1-\alpha) + 2 (1-\alpha)^2 
    + \order\left((1 - \alpha)^3 \right)
\end{equation}
while, for shock-avoiding slicings, an expansion of the function (\ref{sa}) results in
\begin{equation} \label{sa_f_expand}
    f(\alpha) = (1 + \kappa) + 2 \kappa (1 - \alpha)
    + 3 \kappa (1 -\alpha)^2 + \order\left((1 - \alpha)^3 \right).
\end{equation}
In particular, for $\kappa = 1$, the first two terms of the two expansions agree, suggesting that, for values of the lapse close to unity, both conditions will lead to similar results. 

It is also instructive to consider algebraic expressions for the lapse function $\alpha$ in the absence of a shift, in which case the Bona-Masso condition (\ref{bona_masso}) reduces to
\begin{equation} \label{bona_masso_no_shift}
    \partial_t \alpha = - \alpha^2 f(\alpha) K.
\end{equation}
Also, for zero shift, the mean curvature $K$ can be written
\begin{equation}
    K = - \frac{1}{2 \alpha} \, \frac{\partial_t \gamma}{\gamma},
\end{equation}
where $\gamma$ is the determinant of the spatial metric $\gamma_{ij}$, so that (\ref{bona_masso_no_shift}) becomes
\begin{equation} \label{bona_masso_no_shift2}
    \frac{2 \, \partial_t \alpha}{\alpha f(\alpha)} = \frac{\partial_t \gamma}{\gamma}.
\end{equation}

Adopting the choice (\ref{1+log}), \eq~(\ref{bona_masso_no_shift2}) can be integrated immediately to yield 
\begin{equation} \label{1_log_alg}
    \alpha(\gamma) = \alpha_0 + \log(\gamma / \gamma_0),
\end{equation}
where $\alpha_0$ and $\gamma_0$ are initial values.  For $\alpha_0 = 1$, the simple form of \eq~(\ref{1_log_alg}) lends this condition its name ``1+log".  We also note that an expansion of (\ref{1_log_alg}) about $\gamma = \gamma_0$ yields
\begin{align} \label{1+log_expansion}
    \alpha(\gamma) = \, & \alpha_0 
    + \frac{1}{\gamma_0} (\gamma - \gamma_0)
    - \frac{1}{2 \gamma_0^2} (\gamma - \gamma_0)^2  \nonumber \\
    & + \frac{1}{3 \gamma_0^3} (\gamma - \gamma_0)^3 
    + \order\left((\gamma - \gamma_0)^4\right).
\end{align}

Adopting, on the other hand, the shock-avoiding choice (\ref{sa}), \eq~(\ref{bona_masso_no_shift2}) can be integrated to yield
\begin{equation} \label{sa_alg}
    \alpha(\gamma) = \left( \frac{\gamma}{\gamma_0} \left(\alpha_0 + \kappa \right) - \kappa \right)^{1/2}.
\end{equation}
An expansion about $\gamma_0$ now takes the form
\begin{align} \label{sa_expansion}
    \alpha(\gamma) = \, & \alpha_0 
    + \frac{1}{2} \left( \frac{\alpha_0 + \kappa}{\alpha_0 \gamma_0} \right) (\gamma - \gamma_0) \nonumber \\
    &  - \frac{1}{8 \alpha_0} \left( \frac{\alpha_0 + \kappa}{\alpha_0 \gamma_0} \right)^2 (\gamma - \gamma_0)^2 \\
    & + \frac{1}{16 \alpha_0^2} \left( \frac{\alpha_0 + \kappa}{\alpha_0 \gamma_0} \right)^3 (\gamma - \gamma_0)^3   
    + \order\left((\gamma - \gamma_0)^4\right). \nonumber 
\end{align}
Assuming $\alpha_0 = 1$ and adopting $\kappa = 1$, we observe that the first three terms in the expansions (\ref{1+log_expansion}) and (\ref{sa_expansion}) are identical, with the first differences appearing in the cubic term (which enters with a factor of 1/3 in the 1+log expansion (\ref{1+log_expansion}) but with a factor of 1/2 in the shock-avoiding expansion (\ref{sa_expansion})).  As above, we may therefore anticipate that, for values of the lapse close to unity, shock-avoiding slices with $\kappa = 1$ share with 1+log slices some of their desirable properties.

On the other hand, if $\alpha_0 = 0$, the expression (\ref{sa_alg}) does not even allow a regular expansion about $\gamma = \gamma_0$.  This suggests that in regions of strong gravitational fields, for example in the vicinity of black holes, shock-avoiding slices may behave quite differently from 1+log slices.  

In Section \ref{sec:examples} we will confirm both expectations in a number of different examples.

\subsection{Numerics}
\label{sec:numerics}

Our numerical code implements the Baumgarte-Shapiro-Shibata-Nakamura (BSSN) \cite{NakOK87,ShiN95,BauS98} formulation of Einstein's field equations in spherical polar coordinates.  The general strategy of our implementation is described in \cite{BauMCM13,BauMM15}; in particular, the code expresses the BSSN equations adopting a reference-metric formalism (see \cite{Bro09}; see also \cite{ShiUF04,BonGGN04,Gou12}) together with a rescaling of all tensorial quantities in order to handle coordinate singularities at the origin and on the axis analytically.  Unless noted otherwise, the current version of the code evaluates spatial derivatives using eighth-order finite differences, with the exception of advective shift terms, which are evaluated with sixth-order one-sided differencing.  The equations of hydrodynamics, also implemented in spherical polar coordinates with the help of a reference-metric approach \cite{MonBM14}, are solved with a Harten-Lax-van Leer-Einfeld (HLLE)  approximate Riemann solver \cite{HarLL83,Ein88}, together with a simple monotonized central difference limiter reconstruction scheme \cite{Van77}.  All fields are evolved in time using a fourth-order Runge-Kutta integrator.

As coordinate conditions in our simulations we impose the Bona-Masso slicing condition (\ref{bona_masso}) together with a version of a ``Gamma-driver" shift condition 
\begin{equation} \label{shift}
    (\partial_t - \beta^j \partial_j) \, \beta^i
    = \mu_S \bar \Lambda^i - \eta \beta^i
\end{equation}
(see, e.g., \cite{AlcBDKPST03,vanMBKC06}), where $\bar \Lambda^i$ play the role of the connection functions $\bar \Gamma^i$ in the reference-metric formulation of the BSSN equations (see \cite{Bro09,BauMCM13}), and where $\mu_S$ and $\eta$ are parameters.  We usually choose both parameters to be constants, but also consider $\mu_S = \alpha^2$ in some cases (see also \cite{ThiBHBR11,StaBBFS12}).

While our code does not assume spherical symmetry or axisymmetry, all calculations in this paper are performed in axisymmetry, and we therefore need only one (interior) grid-point to resolve the azimuthal angle $\varphi$, $N_\varphi = 1$.  For simulations in spherical symmetry our code requires a minimum of $N_\theta = 2$ interior grid-points for the polar angle $\theta$.  In the absence of spherical symmetry we still assume equatorial symmetry, and use $N_\theta$ uniformly distributed, cell-centered grid-points to cover one hemisphere between $\theta = 0$ and $\theta = \pi/2$.  The radial grid extends from $r = 0$ to $r = r_{\rm out}$ and is constructed from a uniform, cell-centered grid in an auxiliary variable $0 \leq x \leq 1$ with the map
\begin{equation} \label{grid}
    r = r_{\rm out} \frac{\sinh(s_r x)}{\sinh(s_r)},
\end{equation}
where $s_r$ is a constant (see \cite{RucEB18}).  For $s_r = 0$ we recover a uniform grid in $r$, while for $s_r > 0$ the grid is nearly uniform in the vicinity of the origin at $r=0$, but becomes approximately logarithmic at large distance from the origin.  At the outer boundary at we implement simple outgoing-wave boundary conditions for the gravitational fields.  For simplicity we assume that all fields travel at the speed of light, i.e.~we ignore the fact that some gauge modes travel at speeds different from the speed of light.

\begin{table}[]
    \centering
    \begin{tabular}{c|c|c|c|c|c|c}
         Figure & $\eta$ & $\mu$ & $r_{\rm out}$ & $N_r$ & $s_r$ & $N_\theta$  \\
         \hline
         \ref{fig:gauge} & -- & -- & $200 \lambda$ & 4048 & 0 & 2 \\
         \ref{fig:schwarzschild_lapse_central},
         \ref{fig:schwarzschild_lapse_profiles} & 0 & 0.75 & $240 M$  & 256 & 10 & 2 \\  
         \ref{fig:schwarzschild_horizon},
         \ref{fig:schwarzschild_hamiltonian} & 0 & 0.75 & $120 M$  & 512 & 4 & 2\\
         \ref{fig:kerr_horizon} & 0 & 0.75 & $120 M$ & 256 & 6 & 12 \\
         \ref{fig:BL_lapse}, \ref{fig:BL_horizons}, \ref{fig:BL_horizon_surfaces} & 0 & 0.75 & $40 M$ & 512 & 4 & 36 \\
         \ref{fig:os_lapse} & $2 / M$ & $\alpha^2$ & $20 M$ & 512 & 0 & 2 \\
         \ref{fig:tov_density} & 0 & 0.75 & $120 K^{1/2}$ & 512 & 6 & 2 \\
    \end{tabular}
    \caption{Summary of the gauge and grid parameters for the numerical results presented in this paper.  }
    \label{tab:parameters}
\end{table}

We list all parameters used for the different simulations presented in this paper in Table \ref{tab:parameters}.

\section{Examples}
\label{sec:examples}

\subsection{Gauge shocks}
\label{sec:gaugeshocks}

\begin{figure}
    \centering
    \includegraphics[width = 0.48 \textwidth]{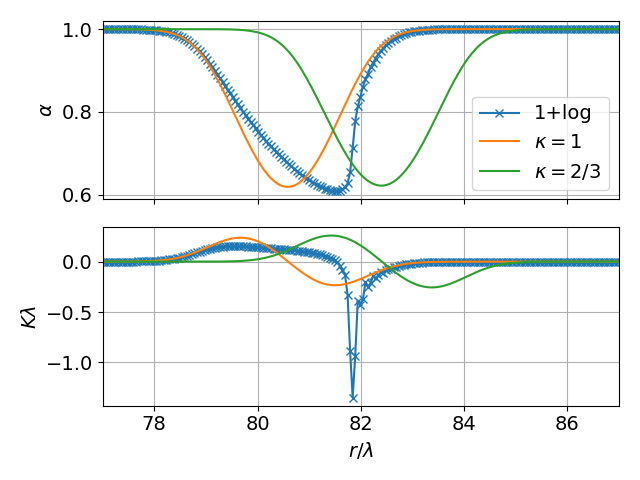}
    \caption{The lapse function $\alpha$ (top panel) and the mean curvature $K$ (bottom panel) for the gauge-pulse test of Section \ref{sec:gaugeshocks}.  Starting with the initial data (\ref{gauge_lapse_init}) we follow the ingoing pulse, and show results for evolutions with 1+log slicing as well as two different versions of the shock-avoiding slicing conditions at time $t = 13.9 \lambda$.  For 1+log slicing we include individual grid-points in order to high-light the steepening of gradients and loss of resolution; the evolutions with the shock-avoiding conditions are performed on the same grid.} 
    \label{fig:gauge}
\end{figure}

As a way of motivating the shock-avoiding slicing conditions (\ref{sa}) we consider a simple gauge-pulse problem previously performed by a number of different authors (e.g.~\cite{Alc97,AlcM98}).  Specifically, we adopt a flat spacetime (expressed in spherical polar coordinates) as initial data.  We also choose the shift to vanish throughout the evolution, but adopt 
\begin{equation} \label{gauge_lapse_init}
    \alpha_0 = 1 - {\mathcal A} \exp\left( - (r - r_c)^2/\lambda^2 \right)
\end{equation}
as the initial lapse, where $\mathcal{A}$, $r_c$ and $\lambda$ are constants.  We then evolve these data with both 1+log and shock-avoiding slicing conditions.  As we will motivate in more detail in Section \ref{sec:schwarzschild}, we adopt two different values of the constant $\kappa$ in (\ref{sa}), namely $\kappa = 1$ and $\kappa = 2/3$.   During the evolution, the initial Gaussian in the lapse function splits, with one pulse moving towards larger and one towards smaller radii.  In the following we focus on the ingoing pulse.

In Fig.~\ref{fig:gauge} we show results for an amplitude ${\mathcal A} = 0.6$ and $r_c = 100 \lambda$ at a coordinate time $t = 13.9 \lambda$.  Even though the initial pulse is well resolved, we see that, for 1+log slicing, one side of the pulse becomes increasingly steep, so that ultimately the solution is no longer adequately resolved.  Simultaneously, the mean curvature $K$ develops a sharp peak, which also cannot be resolved sufficiently well.  In evolutions with the shock-avoiding slicing conditions, on the other hand, neither one of these pathologies develop, and the calculation proceeds without problems -- clearly motivating a study of the properties of shock-avoiding slicing conditions.  

Also note that, for the shock-avoiding slicing condition with $\kappa = 1$, results are very similar to those for 1+log slicing in regions where the lapse is close to unity, as we would expect from the discussion in Section \ref{sec:general}.  For shock-avoiding slices with $\kappa = 2/3$, on the other hand, this similarity disappears. 

\subsection{Schwarzschild spacetimes}
\label{sec:schwarzschild}

\begin{figure}
    \centering
    \includegraphics[width = 0.48 \textwidth]{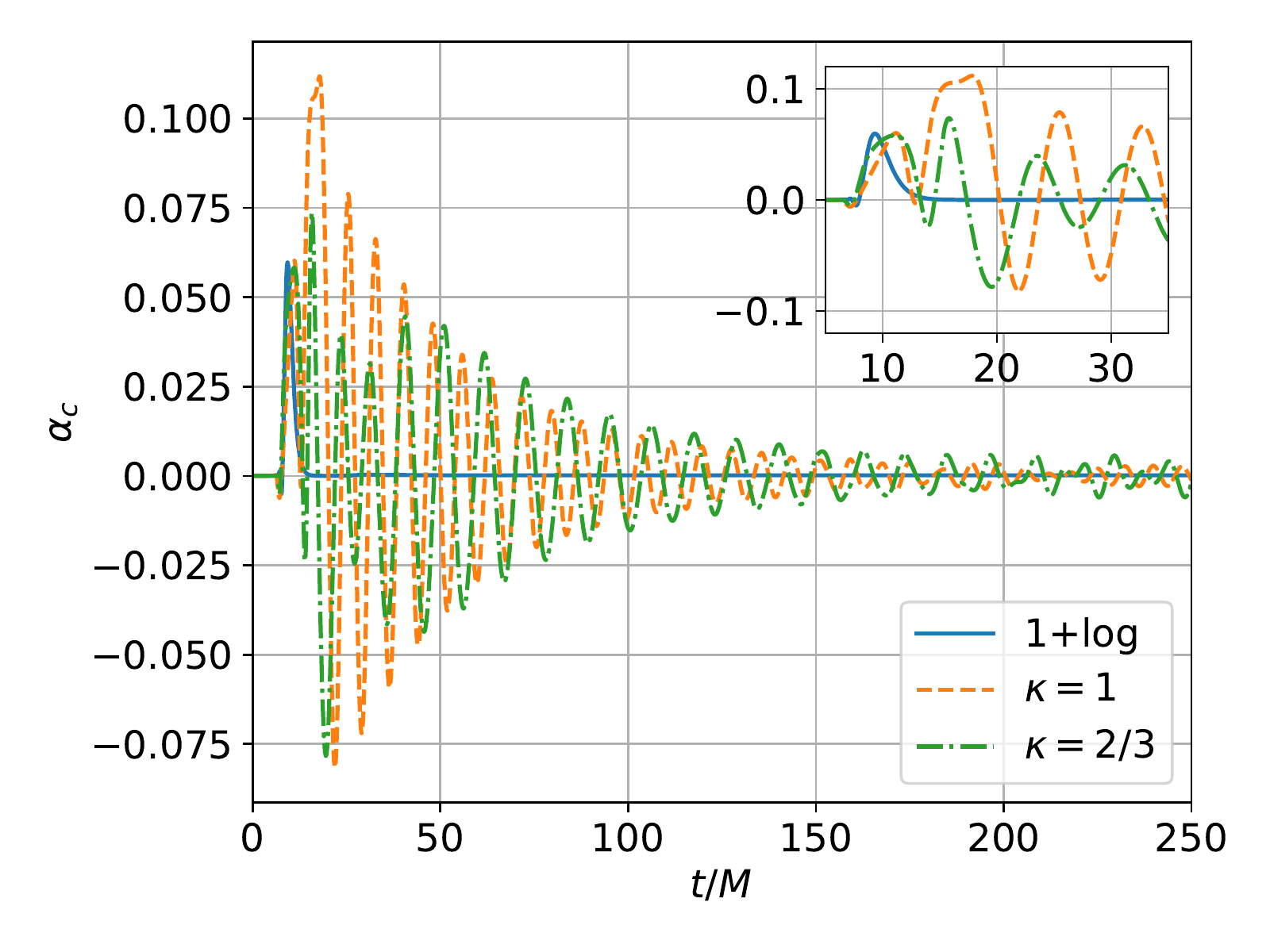}
    \caption{Central values of the lapse function $\alpha$ as a function of coordinate time $t / M$ for the Schwarzschild simulations of Section \ref{sec:schwarzschild}.  We compare results for 1+log slicing with those for shock-avoiding slicings with both $\kappa = 1$ and $\kappa = 2/3$.  Note that, in the transition from a wormhole geometry to a trumpet geometry, 1+log slicing settles down very quickly, while the shock-avoiding slices perform damped oscillations for hundreds of timescales $M$.}
    \label{fig:schwarzschild_lapse_central}
\end{figure}

\begin{figure}
    \centering
    \includegraphics[width = 0.48 \textwidth]{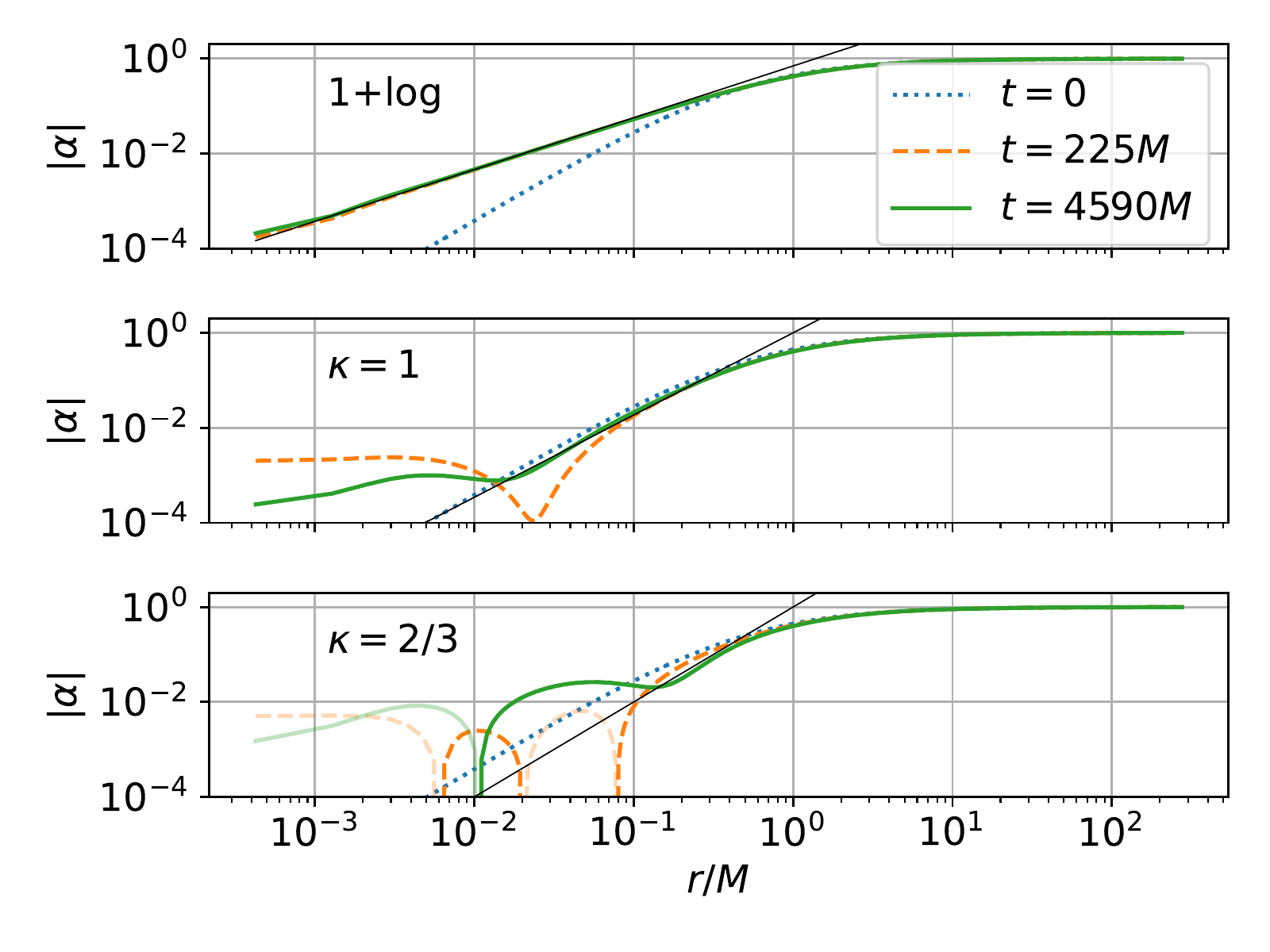}
    \caption{Profiles of the absolute values of the lapse $\alpha$ as a function of isotropic radius $r$ at different instants of time, for 1+log slicing (top panel) as well as shock-avoiding slicing with $\kappa = 1$ (middle panel) and $\kappa = 2/3$ (bottom panel).  Segments of the lines that are faded indicate that the lapse is negative.  The solid black lines show the expected power-law behavior $\alpha \propto r^\delta$ of the respective equilibrium solutions in the vicinity of the puncture $r = 0$, with $\delta \simeq 1.091$ for the 1+log slices (see \cite{Bru09}) and $\delta = \sqrt{3}$ and $\delta = 2$, respectively, for the shock-avoiding slices with $\kappa = 1$ and $\kappa = 2/3$ (see \cite{BaudeO22}.)}  
    \label{fig:schwarzschild_lapse_profiles}
\end{figure}

As a first test of shock-avoiding slicing conditions in curved spacetimes we consider single, non-rotating black holes, i.e.~we perform simulations of Schwarzschild spacetimes.  As initial data we adopt so-called ``wormhole" data, i.e.~the Schwarzschild geometry on a slice of constant Schwarzschild time expressed in isotropic coordinates.  In particular, the initial conformal factor $\psi$ is given by
\begin{equation}
    \psi_0 = 1 + \frac{M}{2r},
\end{equation}
where $r$ is the isotropic radius.  As initial data for the lapse we choose a ``pre-collapsed" lapse $\alpha_0 = \psi_0^{-2}$, so that $\alpha_0 \propto r^2$ in the vicinity of the black-hole puncture at $r = 0$ initially.  We also set the shift vector to zero at the initial time, $\beta^i_0 = 0$.  For a suitable choice of the Bona-Masso function $f(\alpha)$, the evolution  then effectively results in a transition from the spatial wormhole geometry to a so-called ``trumpet" geometry (see \cite{HanHPBM07}).  For 1+log slices, with $f(\alpha)$ given by (\ref{1+log}), this transition has been studied by many authors (e.g.~\cite{Bro08,HanHOBO08}); here we compare 1+log simulations of this transition with simulations adopting shock-avoiding slicing conditions.

Specifically, we compare shock-avoiding slicing conditions (\ref{sa}) for two different values of $\kappa$, namely $\kappa = 1$ and $\kappa = 2/3$.  The former value was adopted by \cite{JimVA21}, and also appears as a natural choice given the discussion in Section \ref{sec:general}.  As shown by \cite{BaudeO22}, this choice leads to an equilibrium solution for which, in isotropic coordiantes, the lapse function behaves as $\alpha \propto r^{\sqrt{3}}$ close to the black-hole puncture.  As an alternative, \cite{BaudeO22} suggested $\kappa = 2/3$, for which the equilibrium solution scales as $\alpha \propto r^2$.

In Fig.~\ref{fig:schwarzschild_lapse_central} we show values of the lapse $\alpha$ at the black-hole puncture $r = 0$ as a function of coordinate time $t$, for 1+log slicing as well as shock-avoiding slicing with $\kappa = 1$ and $\kappa = 2/3$.  For all three slicing conditions, the lapse remains close to zero until about $t \simeq 8M$, before the transition from the wormhole geometry to the trumpet geometry affects the center.  For 1+log slicing, the central lapse again settles down to values close to zero very quickly, after times around $t \simeq 15 M$, while for both choices of $\kappa$ the shock-avoiding slicing conditions lead to a damped oscillation of the central lapse that lasts for hundreds of $M$. 

This behavior can also be seen in Fig.~\ref{fig:schwarzschild_lapse_profiles}, where we show profiles of the absolute values of the lapse as a function of isotropic radius for selected instants of time.  For 1+log slicing, the lapse settles down and assumes its expected power-law scaling of $\alpha \propto r^{1.091}$ close to the puncture (see \cite{Bru09}) very quickly.  For both versions of the shock-avoiding slices, however, the lapse performs oscillations about the equilibrium solution that are damped only rather weakly, so that the expected power-law behavior emerges only after hundreds of $M$.  

\begin{figure}
    \centering
    \includegraphics[width = 0.48 \textwidth]{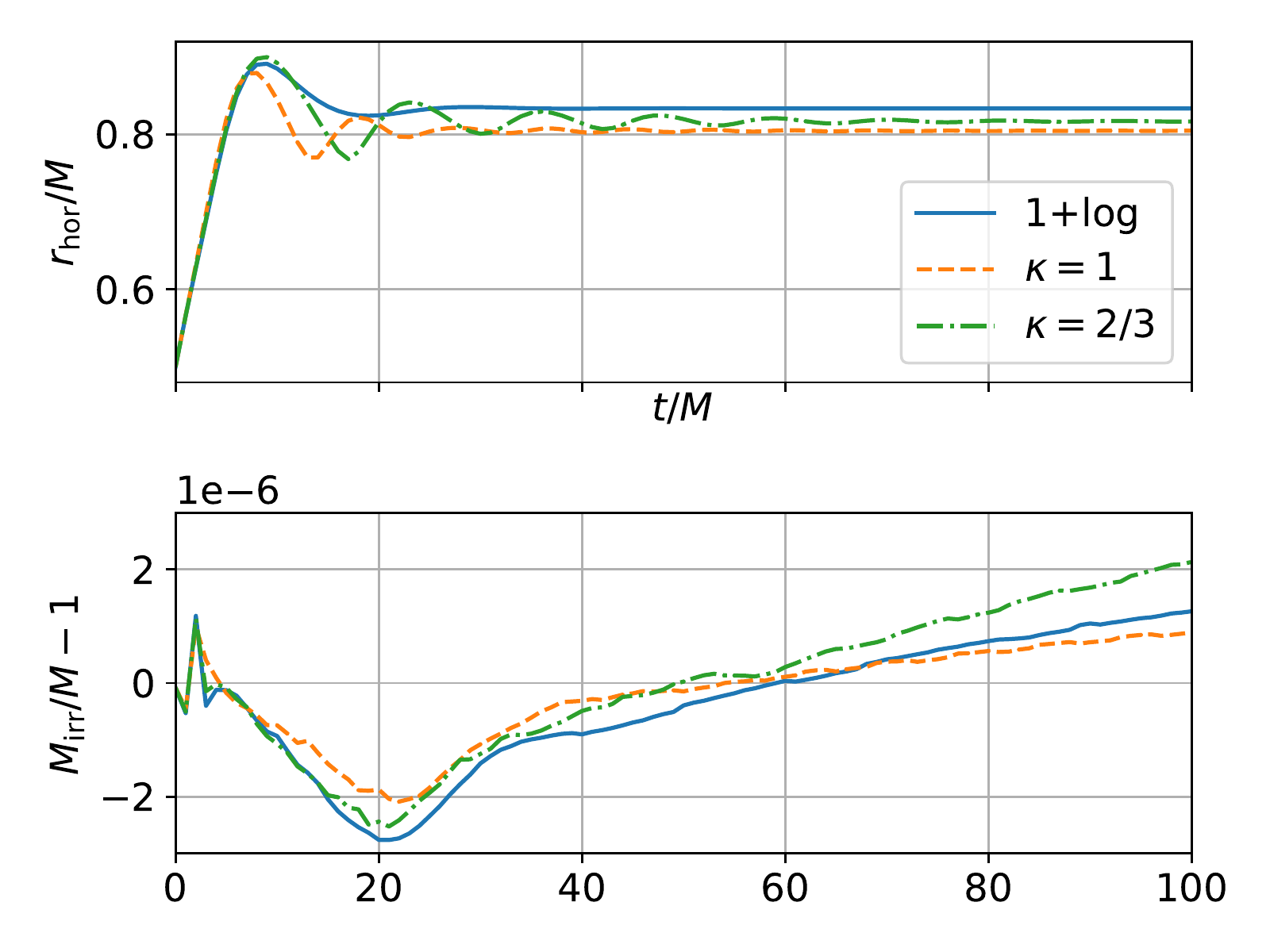}
    \caption{Values of the apparent horizon's coordinate radius (top panel) and mass (bottom panel) for Schwarzschild spacetimes.  For all three slicing conditions the horizon's coordinate location changes as the spatial geometry changes from a wormhole to a trumpet geometry.  For 1+log slicing, however, the horizon settles down to the new location rather quickly, while for the shock-avoiding slicings the horizon oscillates around the new equilibrium for a longer time.  However, neither the overall transition nor the oscillations affect the horizon mass, which remains very close to its initial value for all three slicing conditions.}  
    \label{fig:schwarzschild_horizon}
\end{figure}

\begin{figure}
    \centering
    \includegraphics[width = 0.48 \textwidth]{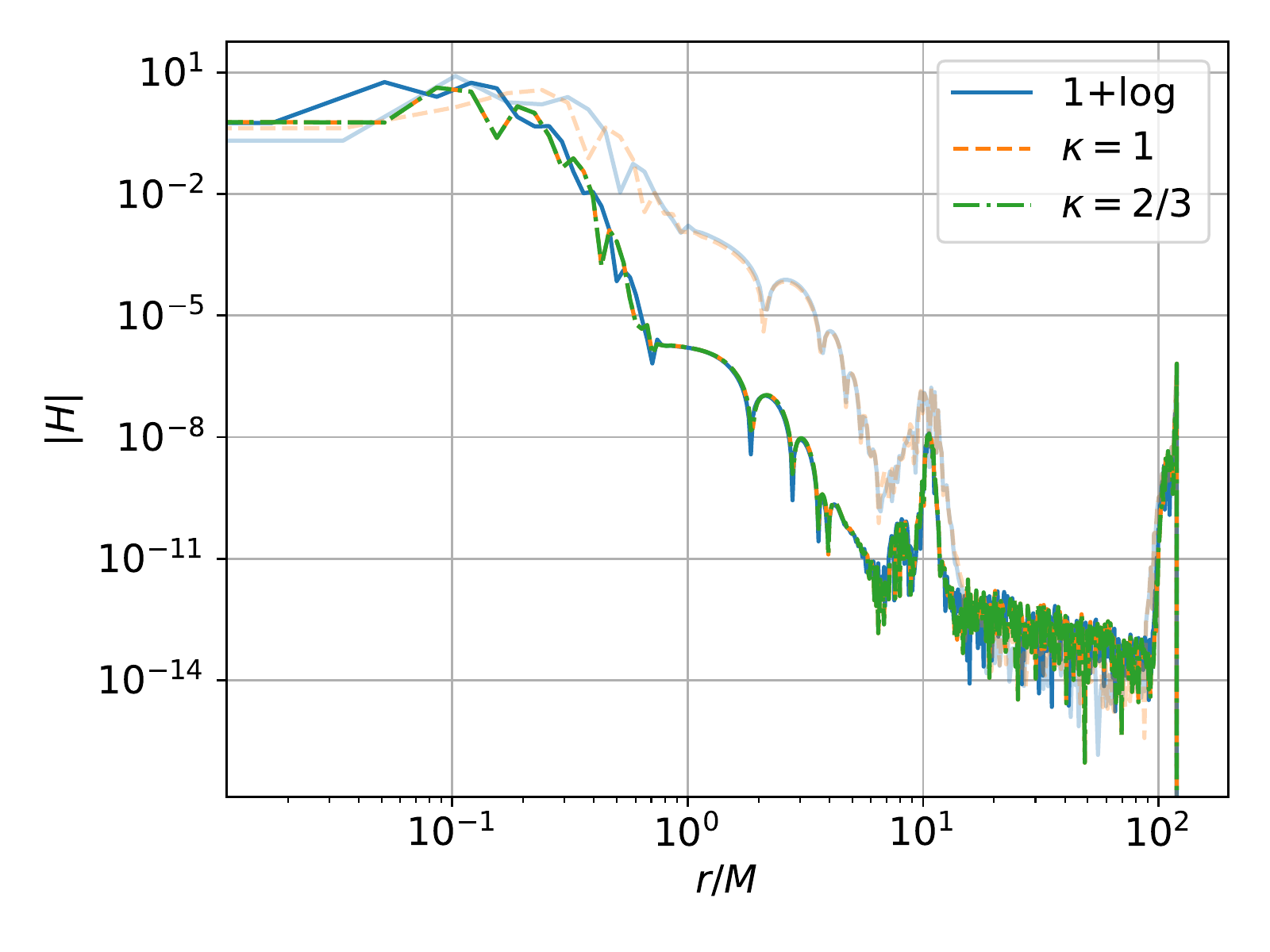}
    \caption{Violations of the Hamiltonian constraint $H$ as a function of isotropic radius $r$ at time $t = 10.8 M$ for Schwarzschild evolution.  The violations are very similar for all three slicing conditions.  In addition for simulations with $N_r = 512$ (see Table \ref{tab:parameters}) we have also included results for $N_r = 256$ as the faded lines.  With the exception of the few innermost grid points in the vicinity of the black-hole puncture, and outer regions that are either affected by noise originating from the outer boundaries or by an apparent floor in the numerical error, the errors converge to fourth order or faster, as expected.}  
    \label{fig:schwarzschild_hamiltonian}
\end{figure}

These oscillations also affect the coordinate location of the apparent horizon, which we show, as a function of coordinate time $t$, in the top panel of Fig.~\ref{fig:schwarzschild_horizon}.  For all three slicing conditions the apparent horizon's location changes as the wormhole geometry changes to a trumpet geometry, but for 1+log slicing this location again settles down to a new equilibrium rather quickly, while for the shock-avoiding slicings the horizon oscillates around the new equilibrium location for a longer time.  The irreducible horizon mass, however, computed from the surface integral
\begin{equation} \label{mass_irr}
    M_{\rm irr} = \left( \frac{\mathcal{A}}{16 \pi} \right)^{1/2}
\end{equation}
where $\mathcal{A}$ is the horizon's proper area, remains very close to its initial value for all three slicing conditions, as shown in the bottom panel of Fig.~\ref{fig:schwarzschild_horizon}.  In fact, the numerical errors are quite similar for all three slicing conditions (at least at times before they are affected by the outer boundary conditions), and also converge quickly with increasing numerical resolution.  This behavior is also shown in Fig.~\ref{fig:schwarzschild_hamiltonian}, where we show violations of the Hamiltonian constraint at a time $t = 10.8 M$.  

Even though the shock-avoiding slices lead to large oscillations in gauge-dependent quantities, with the lapse function taking negative values in some regions of spacetime, it is remarkable that gauge-independent quantities do not seem to be affected by significantly larger errors than those computed with 1+log slices.  While, in evolutions with the shock-avoiding slicing conditions, some quantities do appear to be more affected by noise originating from the outer boundaries than in those with 1+log slicing, and while specific results will of course depend on resolution and the specific of the implementation, we have been able to evolve Schwarzschild black holes to late times with all three slicing conditions. 

\subsection{Kerr spacetimes}
\label{sec:kerr}

\begin{figure}
    \centering
    \includegraphics[width = 0.48 \textwidth]{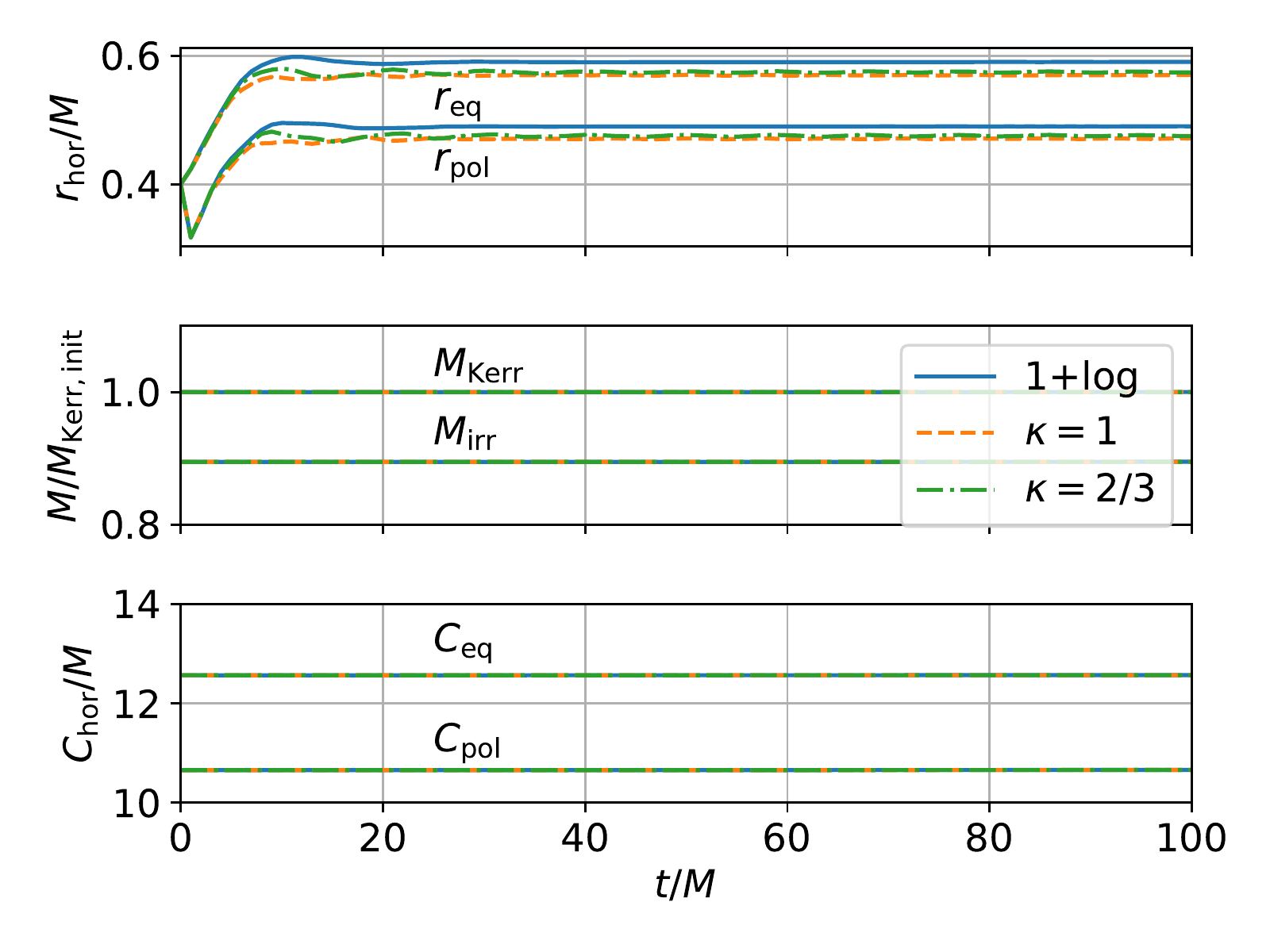}
    \caption{Values of the horizon's coordinate locations (top panel) horizon masses (middle panel) and proper circumferences (bottom panel) for Kerr spacetimes.  We show the coordinate locations of both the equator, $r_{\rm eq}$, and the pole ,$r_{\rm pole}$, the black hole's irreducible mass $M_{\rm irr}$ (see \eq~(\ref{mass_irr})) and Kerr mass $M_{\rm Kerr}$ (see (\ref{mass_kerr})), as well as proper circumferences along the equator, $C_{\rm eq}$, and through the poles, $C_{\rm pol}$.}  
    \label{fig:kerr_horizon}
\end{figure}

We next consider simulations of rotating black holes, i.e.~Kerr spacetimes.  Specifically, we adopt initial data in the coordinate system suggested by \cite{LiuES09} with a moderate spin of $a = 0.8 M$, where $M = M_{\rm Kerr}$ is the black hole's gravitational mass.  As initial data for the lapse we again adopt a pre-collapsed lapse with $\alpha_0 = \psi_0^{-2}$, but as initial data for the shift vector we adopt the values given by the analytical spacetime solution.  

Perhaps not surprisingly, our results for Kerr black holes are, qualitatively, very similar to those for Schwarzschild black holes.  Evolving with 1+log slicing, the lapse function settles down to a new equilibrium after a coordinate time of about $20 M$, while, evolving with the shock-avoiding slicing conditions, the lapse function performs oscillations about the new equilibrium for significantly longer time.  

These oscillations are also reflected in oscillations of the coordinate location of the apparent horizon, which we show in the top panel of Fig.~\ref{fig:kerr_horizon}.  Specifically, we show both the location of the pole ($r_{\rm pol}$) and that of the equator ($r_{\rm eq}$).  Despite these oscillations in gauge-dependent quantities, gauge-invariant quantities again behave very similarly for all three slicing conditions.  To demonstrate this, we show in the middle panel of Fig.~\ref{fig:kerr_horizon} both the irreducible mass (\ref{mass_irr}) as well as the black hole's gravitational, or ``Kerr mass",
\begin{equation} \label{mass_kerr}
    M_{\rm Kerr} = M_{\rm irr} \left( 1 + \frac{1}{4} \, \left( \frac{J}{M_{\rm irr}^2} \right)^2 \right)^{1/2},
\end{equation}
where we compute the black hole's angular momentum $J$ from a surface integral over the horizon.  Finally, in the bottom panel of Fig.~\ref{fig:kerr_horizon} we show proper circumferences of the black hole horizon, both the circumference around the equator, $C_{\rm eq}$, as well as the circumference through the pole along lines of constant polar angle, $C_{\rm pol}$.  Results for these gauge-invariant quantities are very similar for all three slicing conditions, and cannot be distinguished in the figure.

\subsection{Head-on collision of two black holes}
\label{sec:head-on}

\begin{figure}
    \centering
    \includegraphics[width = 0.48 \textwidth]{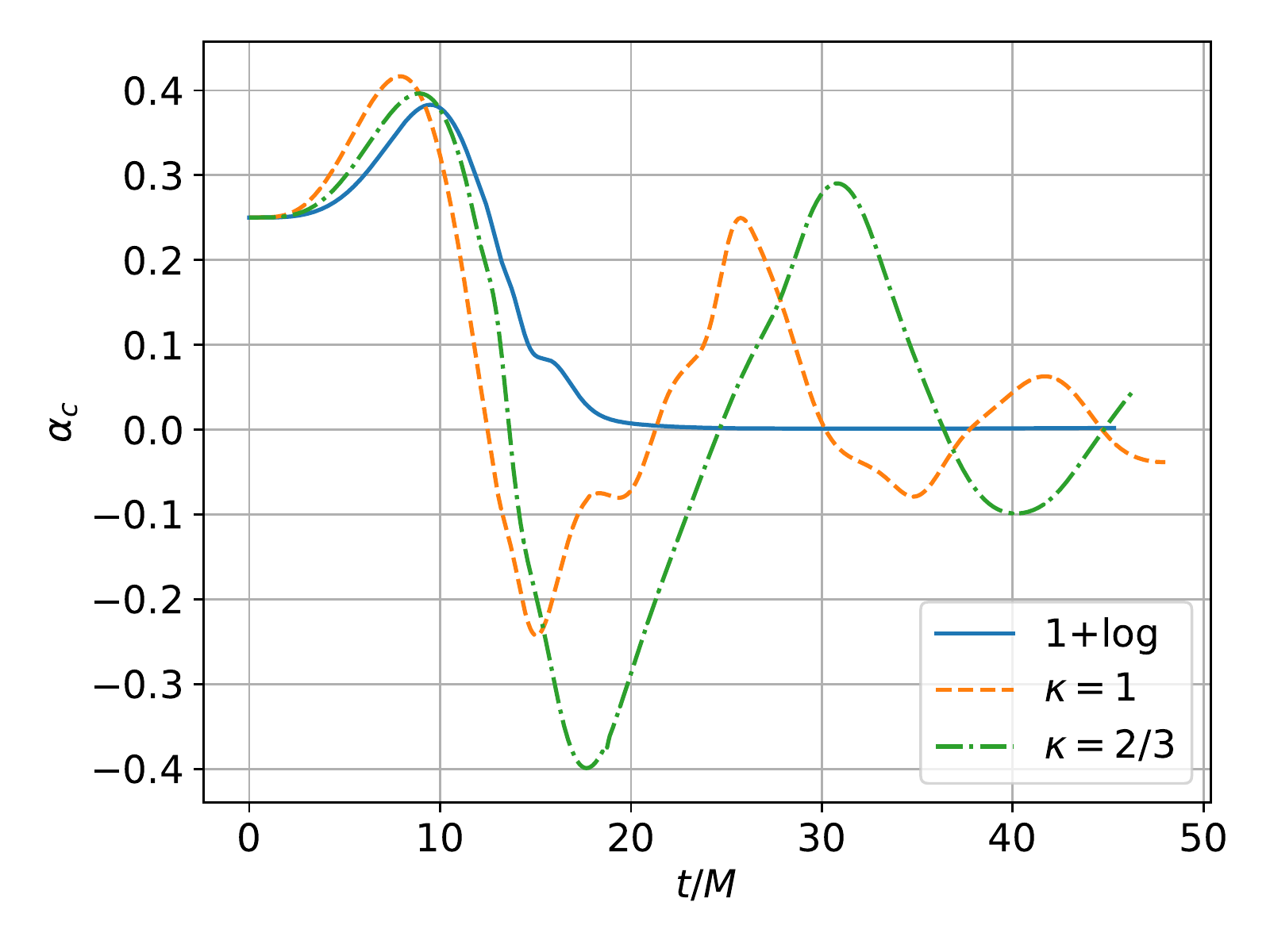}
    \caption{Values of the lapse function at the origin $r = 0$ for the head-on collision of two black holes.}  
    \label{fig:BL_lapse}
\end{figure}

\begin{figure}
    \centering
    \includegraphics[width = 0.48 \textwidth]{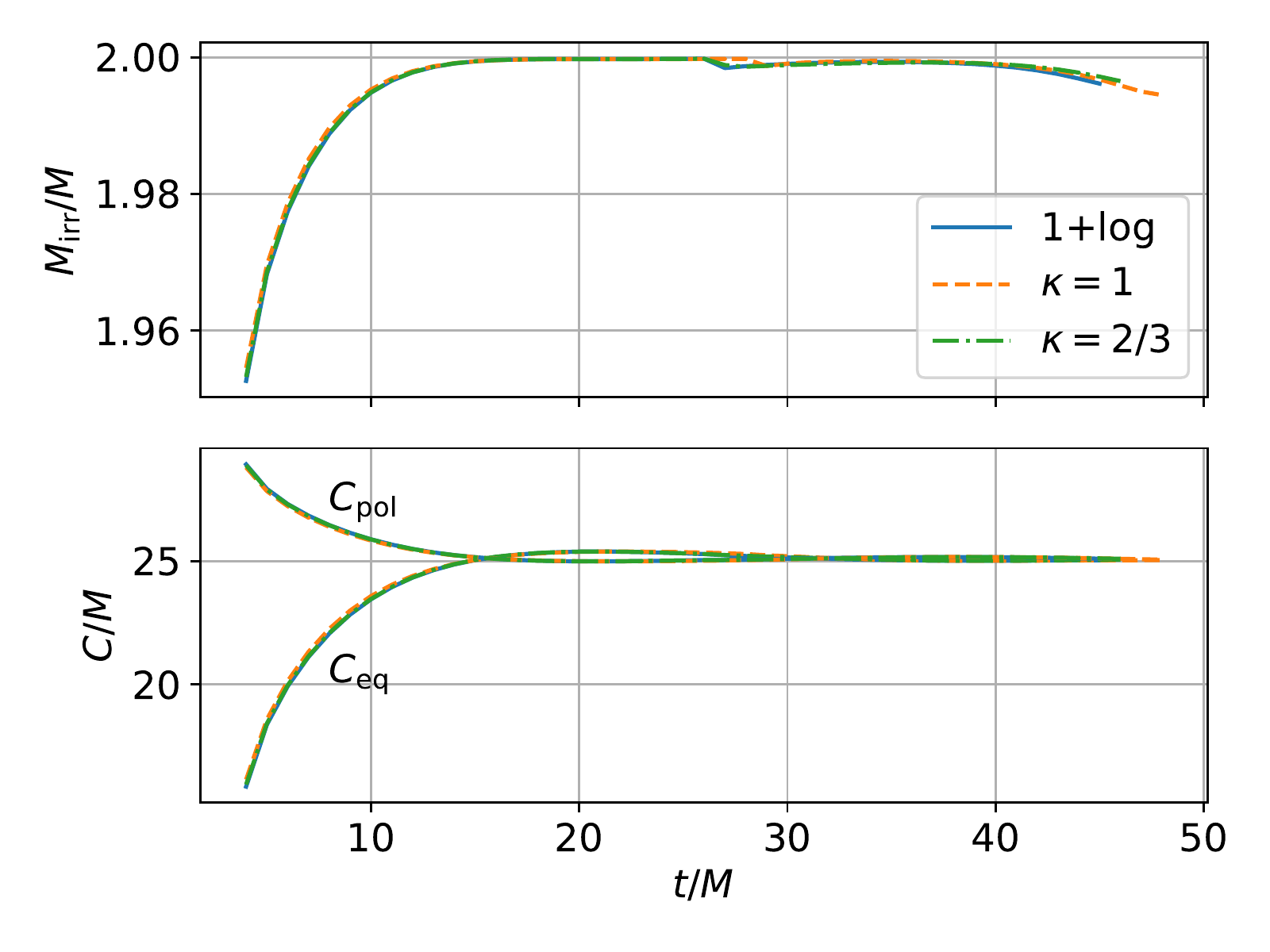}
    \caption{Mass (top panel) and proper circumferences (bottom panel) of the joint horizon that forms in the head-on collision of two black holes.  After $t \simeq 28 M$, the horizon mass in particular is affected by numerical noise originating from the outer boundary, which is located at $r_{\rm out} = 40 M$ in these simulations.  All quantities agree quite well for the different slicing conditions despite the fact that we plot them against coordinate time.} 
    \label{fig:BL_horizons}
\end{figure}

As an example of a truly dynamical spacetime we consider the head-on collision of two black holes. As initial data we adopt ``Brill-Lindquist" \cite{BriL63} data, meaning that the initial slice is conformally flat, time-symmetric, and that the conformal factor is given by
\begin{equation}
    \psi = 1 + \frac{\mathcal{M}_1}{r_1} + \frac{\mathcal{M}_2}{r_2},
\end{equation}
where $r_1$ and $r_2$ measure the coordinate distances from the two black holes.  For our simulations here we adopt $\mathcal{M}_1 = \mathcal{M}_2 = M$, and place the two black holes at locations $z = \pm \mathcal{M}$ on the $z$-axis, so that the problem can be performed with both axisymmetry and equatorial symmetry.  We again choose a pre-collapsed lapse initially (meaning that $\alpha_0 = 0.25$ at the origin), together with vanishing shift.

\begin{figure}
    \centering
    \includegraphics[width = 0.48 \textwidth]{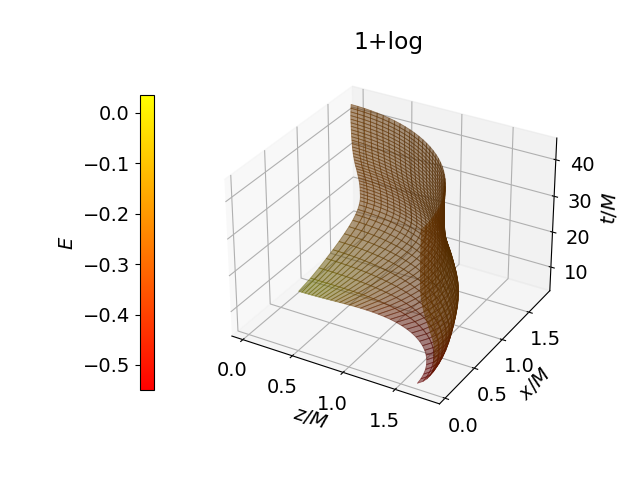}
    
    \includegraphics[width = 0.48 \textwidth]{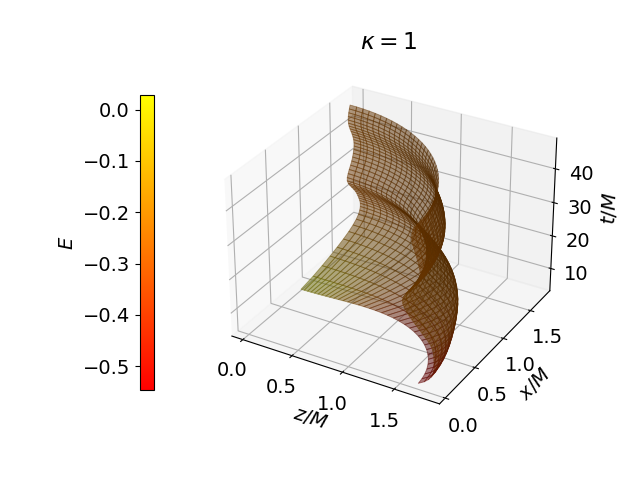}
        
    \includegraphics[width = 0.48 \textwidth]{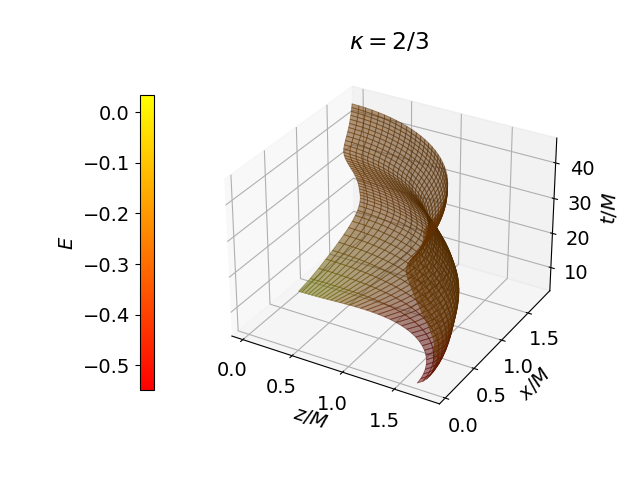}
    \caption{Coordinate locations of the horizon surfaces of head-on collisions, for 1+log slicing (top panel) as well as shock-avoiding slicing conditions with $\kappa = 1$ (middle panel) and $\kappa = 2/3$ (bottom panel).  The horizon surfaces are color-coded by the local tendicity $E$, i.e.~the contraction of the electric part of the Weyl tensor with normal on the horizon.} 
    \label{fig:BL_horizon_surfaces}
\end{figure}

We start by showing, in Fig.~\ref{fig:BL_lapse}, values of the lapse function $\alpha$ at the origin $r = 0$ of the coordinate system.  Once the two black holes have merged, the lapse again settles down to a value close to zero very quickly for 1+log slicing, while the shock-avoiding slicing conditions lead to oscillations.

We first detect a joint horizon at coordinate time close to $t = 4 M$ for all three slicing conditions.  In the top panel of Fig.~\ref{fig:BL_horizons} we show the mass of this joint horizon, which grows initially to the expected value of just below $2M$.  After $t \simeq 25 M$ we observe small deviations, which are caused by noise originating from the outer boundary at $r_{\rm out} = 40 M$. In the bottom panel we show the horizon's proper equatorial and polar circumferences.  Since the merged black hole is distorted initially, these two circumferences are quite different when the joint horizon first forms, but they quickly approach each other as the black hole relaxes to a spherical Schwarzschild black hole.

In Fig.~\ref{fig:BL_horizon_surfaces} we show the coordinate locations of the horizons.  In all three cases the initial horizon is quite distorted, as one might expect, but quickly settles down to a spherical shape.  However, when evolved with the shock-avoiding slicing conditions, the radius performs oscillations similar to those that we have previously observed for Schwarzschild and Kerr black holes.  In Fig.~\ref{fig:BL_horizon_surfaces}, the surfaces are color-coded by the local tendicity $E$, i.e.~the contraction of the electric part of the Weyl tensor with the normal on the apparent horizon (see \cite{Oweetal11}).   While this tendicity takes quite different values at pole and equator at early times it approaches $E = - 1/(2 M_{\rm irr})^2$ everywhere at late times, which is the analytical value for a Schwarzschild black hole.

\subsection{Oppenheimer-Snyder collapse}
\label{sec:os}

\begin{figure}
    \centering
    \includegraphics[width = 0.48 \textwidth]{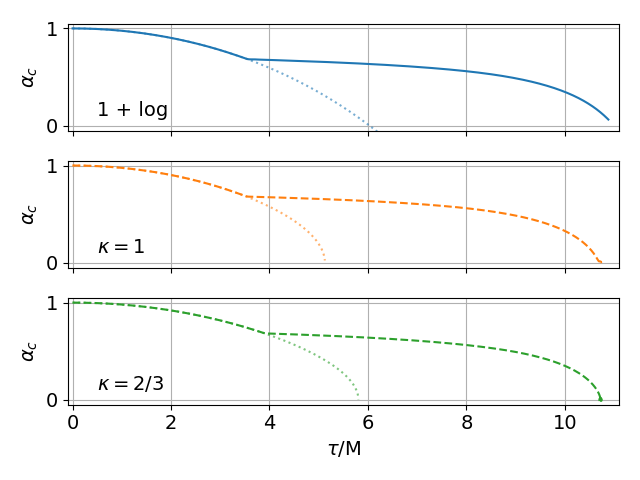}
    \caption{The lapse function $\alpha$ at the center of a collapsing dust sphere as a function of proper time $\tau$. The faint dotted lines represent the solutions (\ref{1_log_alg}) and (\ref{sa_alg}) with $\gamma$ given by (\ref{os_gamma}).  (Compare Fig.~2 in \cite{StaBBFS12}.)}
    \label{fig:os_lapse}
\end{figure}

As a first example that includes matter sources we consider Oppenheimer-Snyder collapse, i.e.~the gravitational collapse of a uniform and spherically symmetric dust ball to a black hole \cite{OppS39}.  Oppenheimer-Snyder collapse in 1+log slicing was analyzed in \cite{StaBBFS12}, and it turns out that much of that analysis can be generalized easily to apply to shock-avoiding slices.

In our simulations we set up the initial data as described in \cite{StaBBFS12}.  In particular, we choose the initial lapse to be one, $\alpha_0 = 1$ and the initial shift to vanish.  As shown in \cite{StaBBFS12}, the lapse will then depend on time only, i.e.~remain spatially constant, in a neighborhood of the center until a gauge mode, originating from the surface of the collapsing star, has reached the center.  Within this neighborhood, slices of constant coordinate time then align with slices of constant proper time, and all quantities remain spatially constant.  Moreover, the arguments presented in \cite{StaBBFS12} show that Eq.~(\ref{1_log_alg}) for 1+log slicing and (\ref{sa_alg}) for shock-avoiding slices, with the determinant $\gamma$ given by
\begin{equation} \label{os_gamma}
    \frac{\gamma}{\gamma_0} = \left( \frac{a}{a_0} \right)^6,
\end{equation}
serve as an analytic expression for the lapse until the gauge mode has reached the center, and as a lower limit afterwards.  In (\ref{os_gamma}), the scale factor $a = a(\tau)$ describing the collapsing dust sphere can be expressed in terms of a parameter $\eta$ as
\begin{subequations}
\begin{align}
    a & = \, \frac{a_0}{2}  \left( 1 + \cos(\eta) \right) \\[1mm]
    \tau & = \, \frac{a_0}{2} \left(\eta + \sin(\eta) \right).
\end{align}
\end{subequations}
In our numerical simulations of this collapse we approximated the dust evolution by solving the equations of relativistic hydrodynamics with the pressure chosen sufficiently small that it does not affect the dynamics. 

In Fig.~\ref{fig:os_lapse} we show results for the lapse $\alpha$ at the center of the collapsing star.  For all three slicing conditions we find excellent agreement with the analytical expressions while the lapse remains spatially constant in a neighborhood of the center.  It is also interesting that for 1+log slicing and for the shock-avoiding slicing condition with $\kappa = 1$ the results agree very well - even plotting them on the same graph it would be very hard to distinguish them until after the gauge mode has reached the center.  This is because, for $\kappa = 1$ and $\alpha_0 = 1$, the first three terms in the expansions (\ref{1+log_expansion}) and (\ref{sa_expansion}) agree, and differences between the two results scale with $(\gamma / \gamma_0)^3$ and are hence very small.

Numerical simulations of Oppenheimer-Snyder collapse are difficult because of the discontinuity of the matter at the stellar surface, which spoils convergence.  In fact, we performed the simulations in this section with fourth-order spatial differencing rather than higher order, since in the domain of dependence of the surface, lower-order schemes performed better than higher-order schemes.  This issue is independent of the slicing condition, of course, and we have not found that the shock-avoiding slices are more affected by this problem than 1+log slices.

\subsection{Neutron stars}
\label{sec:tov}

As a final example we consider the evolution of non-rotating, spherically symmetric neutron stars.  To construct initial data we solve the Tolman-Oppenheimer-Volkoff (TOV) equations \cite{Tol39,OppV39} for a gas with polytropic equation of state
\begin{equation}
    P = K \rho_0^\Gamma,
\end{equation}
and then evolve these data using an ideal gas law
\begin{equation}
    P = (\Gamma - 1) \, \epsilon \rho_0.
\end{equation}
In the above equations $P$ is the pressure, $\rho_0$ the rest-mass density, and $\epsilon$ the specific internal energy density, in terms of which the total energy density is given by $\rho = \rho_0 ( 1 + \epsilon)$.  We have also introduced the adiabatic index $\Gamma$ as well as the polytropic constant $K$.  In the following we adopt $\Gamma = 2$, in which case $K^{1/2}$ has units of length and forms a natural length scale.

\begin{figure}[t]
    \centering
    \includegraphics[width = 0.48 \textwidth]{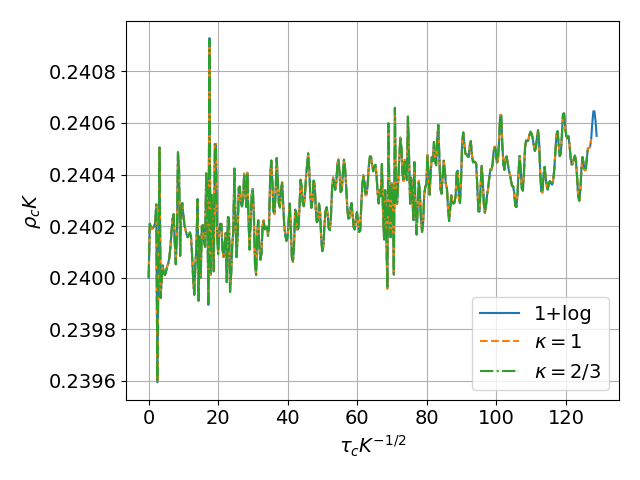}
    \caption{The central energy density $\rho_c$ as a function of central proper time $\tau$ for the neutron star of Section \ref{sec:tov}.  }
    \label{fig:tov_density}
\end{figure}

For our specific initial data we chose a model with central densities $\rho_{0c} = 0.2 K^{-1}$ and $\rho_c = 0.24 K^{-1}$, for which the solution of the TOV equation yields a star with rest mass $M_0 = 0.176 K^{1/2}$, gravitational mass $M = 0.157 K^{1/2}$, areal radius $R = 0.866 K^{1/2}$ and isotropic radius $r = 0.699 K^{1/2}$.  The maximum allowed mass for a $\Gamma = 2$ polytrope is $M_0^{\rm max} = 0.180 K^{1/2}$ and $M^{\rm max} = 0.164 K^{1/2}$.

In Fig.~\ref{fig:tov_density} we show the energy density $\rho$ at the center of the star as function of proper time for evolutions with all three slicing conditions.  The calculations all agree very well, and show only small departures from the initial value.

\section{Summary}
\label{sec:summary}

The 1+log slicing condition \cite{BonMSS95} is among the most successful slicing conditions in numerical relativity and has been adopted in numerous calculations, especially for simulations of compact binaries.  In some applications, however, 1+log slicing leads to gauge ``shocks" in which the lapse function develops discontinuities (e.g.~\cite{Alc97,AlcM98,Hiletal13,JimVA21}).  Alcubierre \cite{Alc97} therefore suggested an alternative choice designed to avoid such gauge shocks.  Even though these ``shock-avoiding" slicing conditions can be shown to perform better than 1+log slicing in some examples, they are also known to have some odd properties -- in particular they allow the lapse function to become negative.

In this paper we compare the performance of 1+log and shock-avoiding slicing conditions for a number of ``text-book" examples, including both vacuum cases and cases with matter, as well as examples with or without spherical symmetry.  We also compare shock-avoiding slicing conditions for two different choices of the free parameter $\kappa$: one with $\kappa = 1$, which was the choice, for example, in \cite{JimVA21}, and one with $\kappa = 2/3$, which has been considered in \cite{BaudeO22}.   Some similar comparisons of evolutions with different Bona-Masso functions $f(\alpha)$, from the perspective of using spectral methods without black-hole excision, are presented in \cite{Oli22}. 

Even though we found that in simulations involving black holes the shock-avoiding slices do indeed allow the lapse to become negative in some regions of the spacetime, this does not appear to affect the stability of the evolution.  In fact, we were able to evolve our examples to just as late times with the shock-avoiding slicing conditions as with 1+log slicing, and numerical errors, at least ignoring those originating from the outer boundaries, are quite similar.

One disadvantage of the shock-avoiding slicing conditions is that they lead to more ``gauge-dynamics" than the 1+log slicing condition: while, for the latter, solutions settle down to time-independent coordinates rather quickly, the former may lead to oscillations about the equilibrium solution that persist for significantly longer.  We also found that, for $\kappa = 2/3$, these oscillations generally appear to be slightly larger and to persist longer.  

While shock-avoiding slicing conditions do allow the lapse function to become negative in regions of the spacetime, and while they may introduce more ``gauge-dynamics" than 1+log slicing, we have found them to perform very similarly in terms of stability and accuracy for all the examples that we considered in this paper.  We therefore believe that Alcubierre's shock-avoiding slicing conditions provide a very viable alternative to 1+log slicing whenever the latter leads to gauge shocks or other pathologies.

\acknowledgments

It is a pleasure to thank Miguel Alcubierre, Carsten Gundlach, and Henrique de Oliveira for many helpful conversations and comments.  This study was supported by the Research Fellowship program at the Mathematisches Forschungsinstitut Oberwolfach in 2022; we greatly appreciate the institute's and its staff's hospitality and support during our stay.  This work was also supported in part by National Science Foundation (NSF) grant PHY-2010394 to Bowdoin College, as well as by the FCT (Portugal) IF Programs IF/00577/2015 and PTDC/MAT-APL/30043/2017 and Project No.\ UIDB/00099/2020.


%

\end{document}